\documentclass[slac_one]{revtex4}
\usepackage{graphicx}
\usepackage{fancyhdr}
\begin{document}

\title{Searches for the Higgs Boson} %% Paper title goes here


\author{M. Herndon for the Babar, CDF and D\O~ collaborations}
\affiliation{University of Wisconsin Madison}

\begin{abstract}
Searches for this Higgs boson are reaching an exciting time.  This proceeding reports the
progress on Standard Model Higgs boson searches at the Tevatron and the prospects for Higgs boson
searches at the Large Hadron Collider.  Also reported are the results of non Standard Model
searches and prospects at the Tevatron, Large Hadron Collide and B factories.  Included in this result are the
first limits on the mass of the Standard Model Higgs boson at high masses beyond the kinematic reach of the
Large Electron Positron collider.
\end{abstract}

\maketitle

\thispagestyle{fancy}

\section{Introduction} % Section title should be in all capitals.
The Higgs boson occupies a unique place in the Standard Model (SM) of particle physics and
in many models of beyond the SM (BSM) physics.  Experimental evidence shows that the strengths
of the electromagnetic (EM) and weak forces are orders of magnitude different at low energies.  The SM proposes
that EM and weak forces are aspects of a single unified electroweak force with similar
coupling strengths governing all the electroweak interactions.  The difference in the strength
of the EM of weak forces results from the massive nature of the W and Z weak bosons compared to the
photon.  In the SM the mechanism of electroweak symmetry breaking that results in the difference in
strength between EM and weak forces at low energy and gives mass to most of the SM particles is known
as the Higgs mechanism.  
This theory predicts a scalar field and an associated scalar boson, 
the Higgs boson.  The existence of this boson
is the primary testable hypothesis of the SM Higgs mechanism.  Discovery of
the SM Higgs boson would be a key element confirming the predictions of the SM.
In addition, many BSM models also predict one or more scalar bosons as part of their
electroweak symmetry breaking mechanisms
and discovery of those bosons would simultaneously allow us to further understand electroweak symmetry
breaking and conclusively prove physics beyond the SM.

In the context of the SM constraints from measuring parameters can be used to
predict the Higgs boson mass.  These constraints primarily come from measurements of the W boson and top
quark masses.  The current fit of all electroweak parameters produced by the  Large Electron Positron 
(LEP) Electroweak Working Group
predicts that the Higgs mass is $84 + 34 - 26$~GeV~\cite{electroweak,notation}.  At 95\% confidence level the 
Higgs boson mass
constrained to be less than $154$~GeV.   These numbers can be compared to the final limit from the
LEP experiments of $114$~GeV at 95\% confidence level (CL)~\cite{lepdirect}.  This indicates that the Higgs boson mass
has a probable value in the area of best sensitivity for the Tevatron experiments.  Also of interest
is that the central value of the predicted mass of the Higgs boson is below the LEP limit indicating
that the Higgs boson might not be of SM origin.

In this proceeding I review the state of experimental searches for the Higgs boson and discuss
prospects for finding the Higgs boson at future colliders.  The results from the Tevatron are
based on an integrated luminosity of up to $3~fb^{-1}$ of $p\bar{p}$ collisions at 1.96 TeV.
Results from the Babar experiment are based on a dedicated run at the $\Upsilon(3S)$ resonance at the PEPII
collider. 
In this review I begin by discussing the experiments and techniques used to search for the Higgs boson.
I then review the status of
searches for BSM Higgs bosons, the status of searches for SM Higgs bosons and conclude with prospects
for Higgs boson searches at future colliders.

\section{Experiments and Techniques}
The field of Higgs boson searches is rich and allows for searches at a number of collider experiments.
I will primarily review searches performed the the Tevatron experiments CDF and D\O~.  The Tevatron is a 
proton anti-proton collider operating at 1.96~GeV.  The Tevatron experiments have recorded $4~fb^{-1}$ of integrated
luminosity of which up to $3fb^{-1}$ has been analyzed for Higgs boson searches.  The CDF and D\O~ experiments
consists of magnetic spectrometers instrumented with tracking and vertex finding detectors surrounded by
electromagnetic calorimeters for measuring the energy of electromagnetic particles such as photons and electrons and 
hadronic calorimeters for measuring the energy of quark jets, which are in turn surrounded by dedicated chambers for detecting
muons.  This design in general allows for the measurement of momentum of all charged particles such as electrons and muons, the energy of photons
and quark jets, and the detection of displaced vertices's characteristic of the decay of b flavored hadrons in
quark jets.  In addition, triggering of interesting events is very important at the high crossing rates of
collider experiments.  The CDF and D\O~ experiments have the ability to trigger on leptons and photons
as well as the energy of jets and the overall missing energy in the transverse plane.  In addition, it is possible to search
for lighter Higgs bosons at the dedicated B physics experiments.  The dedicated B physics experiments have been
primarily running at an energy to produce the $\Upsilon (4S)$ resonance, but recently the Babar experiment has 
collected smaller datasets at the 
$\Upsilon (3S)$ and $\Upsilon (2S)$ resonances, which can be used to search for light Higgs bosons.  The design of the B physics
experiments is similar to the hadron collider experiments, though the typical low track multiplicities of the
B physics environment has allowed the detectors to be optimized for higher efficiency and strong particle identification
capabilities.  Finally, the Large Hadron Collider (LHC) experiments, CMS and ATLAS, will start collecting data soon.  
The LHC is a 14TeV proton proton collider that will possibly achieve luminosities two orders of magnitude
higher than those achieved at the Tevatron.  The
CMS and ATLAS experiments are similar to the Tevatron experiments though generally have physics object detection
capabilities over a larger solid angle.  The strong detectors, coupled with the high luminosity and larger cross
sections for Higgs boson production makes it likely that the LHC can detect a Higgs boson at any reasonable mass.

Successful Higgs boson searches require excellent detector performance in identifying a variety of physics objects.
Signatures that occur in Higgs production and decay include, isolated electrons, muons and photons; tau leptons
which can decay to lighter leptons or hadrons; light quark and b quark jets; and missing transverse energy
from neutrinos or other undetected particles.  

The first step to detecting Higgs boson events is identifying
events to be saved for later analysis using triggering systems.  A primary set of triggers for many Higgs boson
searches are high transverse momentum electron and muon triggers.  These triggers generally detect electrons
using the electromagnetic calorimeter and muons using dedicated muon chambers.  The CDF detector additionally uses
their drift chamber to measure the momentum of the muons while the D\O~ has an array of toroidal magnets that allows
the momentum of the muons to be directly measured in the muon detectors.   The CDF detector has considerable gaps
in solid angle coverage for muons and to a lessor degree electrons.  The dedicated charged lepton triggers are
supplemented by triggers for missing transverse energy and jets as measured in the calorimeters, 
which are effective since if a charged leptons is not
detected then it will leave a signature of apparent missing transverse energy.  For topologies with no charged
leptons the CDF and D\O~ detectors rely purely on the missing energy and jet triggers.  In addition, the experiments
use dedicated tau lepton triggers, which search for narrow jets identified in the calorimeter accompanied by
a small number of charged tracks.

Once events have been triggered and saved for further analysis more sophisticated algorithms can be applied
to identifying physics objects of interest.  The purest physics objects for Higgs searches involve high transverse momentum electrons and muons.
These leptons are identified as 
well measured tracks with consistent information from the dedicated leptons identification systems.  At the D\O~
detector this strategy allows identification of electrons and muons over a large range of solid angle while at CDF
these methods are supplemented by identifying high pt tracks that are either isolated in terms of calorimeter
energy along the path of the tracks, indicating muons, or isolated in terms of other tracks around the candidate
charged leptons, which can identify muons, electrons and single prong hadronic tau lepton decays.
Dedicated tau lepton detection algorithms looks for narrow jets of particles with one or three charged tracks
with only a small amount of energy deposited in an annulus around the tau flight direction.  Additionally the tau
momentum measurement is improved by using both the charged track momentums and the energy in the electromagnetic
calorimeters to include energy for neutral pions.  Furthermore, this information can all be as input to a Neural
Network (NN) to maximize the tau identification performance.

Algorithms for
photons primarily require that there is a well measured electromagnetic energy deposit and no corresponding tracks, but
more advanced algorithms using information from highly segmented presampling or shower maximum calorimeter components
can be applied to reduce large backgrounds from QCD sources such as neutral pions.   Again , this information can all of included in a NN
to maximize photon identification performance.

Quark jets are identified as
energy clusters in the electromagnetic and hadronic calorimeters by summing the energy in a cone around the highest
energy deposit.  The energy resolution can be further improved by using tracking information to measure the
charged track component of the jets.  
Quark jets from b quarks are of particular interest in Higgs boson searches.
Identification of b quark jets is known as b tagging.
The D\O~ b tagging algorithm combines information in jets such as the decay length
of secondary vertices's, displaced impact parameters of charged tracks, and leptonic decay information
in a NN designed to give a continuous output that indicates the likelihood that
a jet was produced by a b quark.  Several operating points can be utilized from the output to
achieve varying levels of efficiency and purity.
The CDF experiment employs a b quark identification algorithm based on well identified
secondary vertices's. 
The invariant mass of the tracks forming the vertex can be used to estimate
the percentage of different quark flavors in the background.
Additionally CDF employs a jet probability algorithm which uses the displaced impact parameters of charged tracks.
CDF does not use leptonic decay information, reserving that information to calibrate the b tagging performance.

Finally missing transverse energy is identified using the vector sum of 
calorimeter energy corrected by the momentum of muons or charged leptons identified as isolated tracks.

\section{BSM Higgs}
Many new physics models predict the existence of one or more Higgs bosons.  In these models the Higgs
boson occupies a similar role in electroweak symmetry breaking as in the SM.  
One subset of new physics
models know as Supersymmetry (SUSY) models has inspired many dedicated searches for BSM Higgs bosons.  In
the minimal SUSY extension of the SM, MSSM, the Higgs sector is expanded to five Higgs bosons.
Three of the five Higgs bosons are neutral with a CP even light Higgs, $h$, a CP even heavy Higgs, $H$, and a
CP odd Higgs boson, $A$.  The two remaining Higgs bosons are the singly the charged $H^+$ and $H^-$.  In 
these models the coupling of the neutral Higgs bosons to $b$ flavored quarks and $\tau^-$ leptons can be
enhance by a factor known as $\tan{\beta}$.  This can enhance both the production of the Higgs bosons,
if there are $b$ quarks in the initial state, and the decay to both $b$ quarks and $\tau^+$ leptons.
In addition, at certain masses either the light or heavy neutral CP even Higgs bosons can be 
degenerate with the neutral CP odd Higgs boson and both can be searched for simultaneously.
This can result in a significant enhancement of the potential sensitivity for observing a Higgs boson
with the production cross section and decay enhanced over SM values by as much as two orders of 
magnitude.

In addition to standard MSSM models, there are many other BSM models that predict enhanced production
of Higgs bosons.  For instance, in fermiophobic Higgs models the couplings to fermions are reduced and
the Higgs boson decays primarily to bosons.  Signals involving photons or W and Z bosons can be distinctive
enhancing the sensitivity to detect Higgs bosons produced in these models.  In addition, there are models
that predicts Higgs bosons below the LEP limits for direct Higgs boson searches, which could be visible
at hadron colliders or B factory experiments.

\subsection{SUSY Higgs Searches}
\begin{figure*}[t]
\centering
\includegraphics[width=135mm]{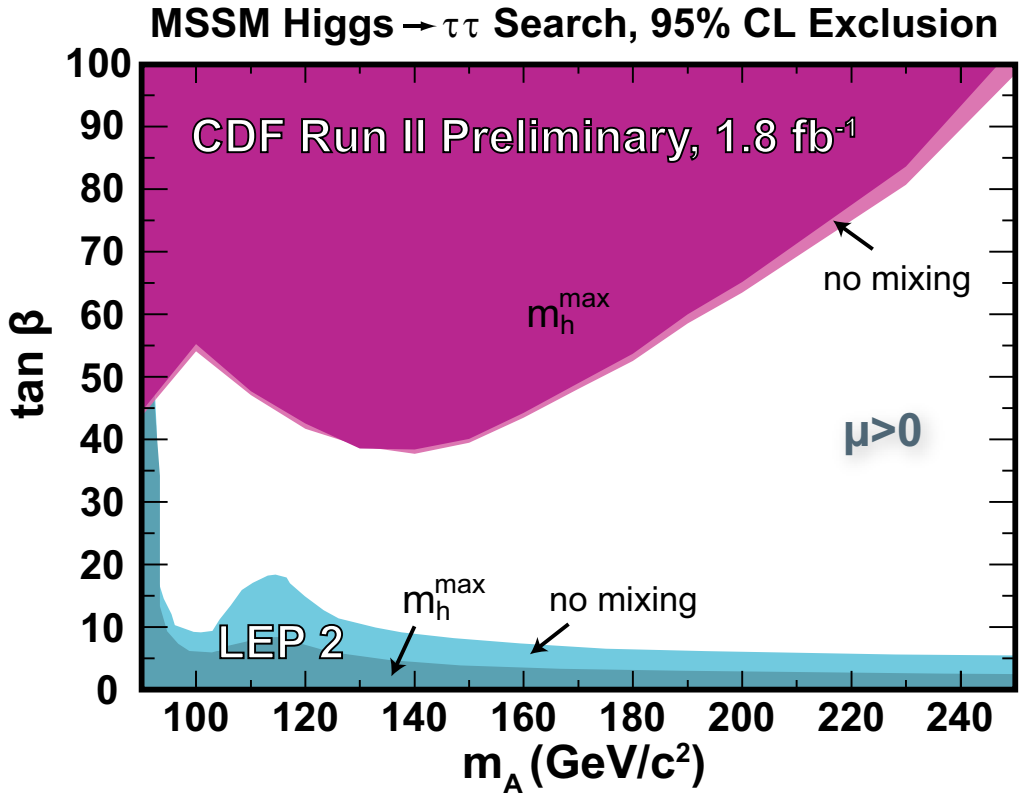}
\caption{Limits as a function of $m_A$ vs $\tan{\beta}$ for the neutral SUSY Higgs $\tau^+\tau^-$ search.} 
\label{susyhtt}
\end{figure*}

The Tevatron experiments CDF and D\O~ have recently performed searches
for neutral SUSY Higgs bosons.  These searches can be used to put limits on the
allowed space in a two dimensional plane of $\tan{\beta}$ and the $m_A$, the mass of the CP odd neutral
Higgs boson.
These searches have been performed in three primary
channels.  The first is an inclusive search for neutral Higgs decays to $\tau^+\tau^-$ pairs.
The key issue of this search is optimizing tau identification efficiency and rejection
of fake taus from jets.  To maximize acceptance the CDF and D\O~ experiments search for tau leptons
in leptonic decays to muons and electrons and hadronic decays and combine them into channels
where the final states are either fully leptonic or one tau decays hadronically.
The purity of the $\tau^+\tau^-$ signature in this mode gives this search competitive
sensitivity.
The performance of the tau finding algorithms is optimized using large samples of W and Z boson
decays involving tau leptons.
The results of the CDF and D\O~ searches are preliminary and use $2~fb^{-1}$  and 
$2.2~fb^{-1}$ of integrated luminosity respectively.  
Both searches place limits on $\tan{\beta}$
to be less than order 50 over a wide range of CP odd Higgs masses.
The world's best limits from the CDF experiment in the space of  $m_A$ vs $\tan{\beta}$ for the $m_{hmax}$~\cite{mhmax} scenario
are shown in figure~\ref{susyhtt}
More details on limits in other scenarios and all the analyses described in this proceeding are given on the Babar, CDF and D\O~ web pages~\cite{cdfd0web}.

The D0 experiment additionally performs a search where the Higgs boson is produced in association with a
b quark.  The Higgs boson decays to a $\tau^+\tau^-$ pair and the b quark is identified as a jet
with properties indicative of the relatively long life or decay patterns of b flavored hadrons using
the b tagging algorithm described above.   Using these tools the D0 performed a search
based on $2.2~fb^{-1}$ that achieves similar sensitivity to the inclusive searches for Higgs decays
to tau pairs.

The CDF and D\O~ experiments also perform a search for Higgs produced in association with b quarks
and decaying to pairs of b quarks.  In this search the experiments identify events with
three or more jets
where at least three of the jets are identified as from b quarks and then search for a peak in the invariant
mass distribution of the two highest transverse energy, $E_T$, jets.
The key issue of this search is understanding the quark content of the background of events with
three or more jets from QCD processes.  Since, the b tagging algorithms employed for these searches
have substantial mis-identification rates for c quark or lighter quark jets, which vary
as a function of jet $E_T$,  understanding this
background is complex.  The two experiments employ different methodologies to achieve this purpose.
The CDF experiment employs a b quark identification algorithm solely based on well identified
secondary vertices's and further employs the invariant mass of the tracks forming the vertex to estimate
the percentage of different quark flavors in the background.
The D\O~ experiment employs their NN tagger at several operating points in order to create a
series of simultaneous equations that can be used to solve for the percentages of quark flavors
using the know efficiencies and mis-identification rates at each of the operating
points.  The sensitivity of these searches is enhanced by the large branching fraction
to b quarks, but simultaneously limited by the large background.  The CDF and D\O~ experiments find limits
of comparable sensitivity to the tau pair channel in scenarios where the b quark coupling is enhanced.

\subsection{Exotic BSM Higgs Searches}

The CDF and D\O~ experiments conduct a number of searches for fermiophobic Higgs where the Higgs boson
will have dominant coupling to bosons.  They search for both direct gluon fusion production of Higgs
bosons where the Higgs decays to two photons and associated production of a W or Z boson with a Higgs boson
where the Higgs boson decays to a pair of oppositely charged W bosons or photons.  
These searches can be used to search for either fermiophobic
or SM Higgs and have been optimized in a model independent way.  The sensitivity of such searches
to fermiophobic Higgs could be substantially enhanced with a specific optimization for those decays modes.

The D\O~ collaboration has performed a search for Higgs bosons decaying to two photons.  The key element in
this search is to reduce the background of QCD jets being misidentified as photons.  The D\O~ search employs
a NN based photon identification that substantially reduces the backgrounds in this mode.  The resulting search
has been benchmarked as a SM Higgs boson search.  The sensitivity of the search expressed as the ability 
to set limits on SM Higgs boson production cross section at 95\% confidence level(CL) is
23 times the SM Higgs boson production cross section for a Higgs boson of 115~GeV. 
This metric of sensitivity will be used throughout to describe the sensitivity of the SM Higgs boson searches.
The sensitivity for fermiophobic
Higgs boson production has not yet reached the expected fermiophobic Higgs production cross sections for Higgs masses above the LEP limits.

Both the CDF and D\O~ collaboration perform a search for associated production of the Higgs boson with a W or Z boson.
In this search the Higgs boson decays to a pair of W bosons and the signature is same sign charged leptons from one
of the associated bosons and one of the W bosons from the Higgs boson decay.  These searches are also benchmarked as
SM Higgs boson searches and achieve sensitivities of 33 and 20 times the SM Higgs boson production cross section
at 160~GeV for the CDF and D\O~ experiments respectively.   Similarly in these modes sensitivity to the cross section for fermiophobic
Higgs production has not yet been achieved.

\subsection{Low Mass Higgs}
the Babar collaboration has performed a search for low mass CP odd Higgs production below a mass of 7.8~GeV.  This
search utilized the $\Upsilon (3S)$ data set to search for the decay 
$\Upsilon (3S) \rightarrow \gamma A_0 \rightarrow \gamma\chi\chi$.  The neutralinos, $\chi$, are not observed so this
results in a signal of a single photon and missing energy consistent with the CP odd Higgs mass.  The Babar
experiment observes an excess of events, $2.6\sigma$, at a mass of $5.2$~GeV.  This result could be further
investigated using data from the $\Upsilon (2S)$ or using data from other experiments.

\section{SM Higgs}
In the SM the Higgs boson can be produced with cross sections of order $10fb^{-1}$ to $1pb^{-1}$ by a variety
of processes including direct gluon fusion to Higgs which the largest cross section, 
associated production with a W or Z boson and vector boson
fusion with two associated quarks.  The Higgs boson decays with greatest probability to the highest mass available
state which leads to dominant decays of on order 90\% to b quarks pairs and 10\% decays to tau lepton pairs 
at masses around 115~GeV and dominant decays to W boson pairs at mass around 160~GeV.  The cross over point when the
b quark pair and W boson pair branching ratios are the same is approximately 135~GeV.   This knowledge of the Higgs
production and decay characteristics leads to a clear strategy for Higgs boson searches at the Tevatron.  At low masses
the combination of the dominant gluon fusion production mechanism and b quark pair decay is overwhelmed by a large
QCD background and instead we look for associated production with W and Z bosons, where the W and Z bosons decay
leptonicaly.  Additionally the tau meson pair decay mode is distinct enough to search for in all production modes.
At high Higgs mass we search for gluon fusion Higgs production with decay to W bosons pairs which decay leptonicaly.  
Additionally, this decay is distinct enough that the Higgs boson events could be detected in all production modes.
In the following subsection I review the various Higgs searches in order of increasing sensitivity.

\subsection{Secondary SM searches}
In order to maximize sensitivity to the production SM Higgs boson the Tevatron experiments search for the Higgs boson
in several decay channels that individually don't have strong sensitivity to the Higgs boson, but collectively in
combination with all the Tevatron SM Higgs boson searches will help achieve the goal of reaching SM sensitivity.
These channels typically have enough sensitivity to detect on order less than one Higgs boson candidate per $fb^1$, but will
contribute to the overall size of a possible Higgs boson signal.  Two such searches have already been discussed; the
searches for Higgs decays to pairs of photons or W bosons.  The remainder of the searches involve decays
to tau leptons or quarks.

The D\O~ collaboration performs a search for associated production of the Higgs boson with a W boson where the Higgs
boson decays to a pair of b quarks and the W boson decays to a tau lepton and a neutrino.  This search is the first
dedicated search for a Higgs boson in this mode and achieves a sensitivity of 42
times the SM production cross section for a Higgs boson of 115~GeV.

The CDF experiment performs a search for associated production of the Higgs boson with a W or Z boson where the Higgs
boson decays to a pair of b quarks and the the vector bosons decays to a pair of quarks.  This search is the first search
in the four jet mode and achieves a sensitivity of 37
times the SM production cross section for a Higgs boson of 115~GeV.

The CDF experiment performs a search for the Higgs boson decaying to a pair of tau leptons.  In this search the Higgs
boson is searched for in association with two jets which can occur in associated production with the 
W or Z boson where the bosons decay to jets, vector boson fusion with two associated quark jets, or gluon fusion where there initial state gluons
radiate gluons which forms jets.  The two tau leptons decay to either hadrons or lighter charged leptons and at least
one of the tau leptons is required to decay to a charged lepton to give a pure enough final state to be distinct from
the background.
This search is particularly interesting since it was the first search to simultaneously consider these three production
mechanisms and it will be an important search mode for low mass Higgs at the LHC.  The sensitivity of this search is 25
times the SM production cross section for a Higgs boson of 115~GeV.  This search adds 5\% sensativity to the overall CDF combination
of SM Higgs searches
demonstrating the importance of searching for the Higgs boson in all viable production and decay modes.

\subsection{$ZH \rightarrow \ell^+\ell^- b\bar{b}$}
The first of the three most highly sensitive low mass
Higgs search modes is the search for associated production of a Higgs boson
with a Z boson where the Z boson decays to a pair of light charged leptons and the Higgs boson decays to a pair
of b quarks.  The unique feature of this decay mode is that it is fully reconstructed making it one of the most pure
Higgs signals available.  Since background is not a primary issue the goal of this search is to maximize
b jet tagging and lepton finding efficiency.  The CDF and D\O~ experiments pursue a strategy of using events where
both b jets are tagged with a loose high efficiency and low purity tag and events where one jet is b tagged
with a tight lower efficiency higher purity tag.  In addition, the CDF experiment uses several categories of
leptons which are identified with relaxed cuts on the information from the dedicated lepton detectors or are
only identified as isolated tracks based on calorimeter isolation.  Also CDF uses the fact that the event
topology should have no missing transverse energy to correct jet energies using a NN algorithm primarily based
on information quantifying whether the observed missing transverse energy is collinear with the jet direction indicating jet
energy underestimation.
Furthermore, both experiments apply an array
of advanced techniques to construct discriminating variables with optimum ability to distinguish signal from background.
For instance, multivariate discriminates such as NNs or boosted decision trees (BDTs) are used and additionally
the matrix element(ME) technique, where the differential matrix elements of the signal and background processes are used
to form an event likelihood that an event with given kinematic properties is signal or background like, is employed.
Using integrated luminosities of 2.3 and 2.4 $fb^{-1}$ 
the $ZH \rightarrow \ell^+\ell^- b\bar{b}$ searches from CDF and D\O~ have the sensitivity to see approximately two Higgs events
or a sensitivity of 11.8 and 12.3 times the SM production cross section for a Higgs boson of 115~GeV respectively.

\subsection{$VH \rightarrow MET b\bar{b}$}
The second of the three most highly sensitive low mass
Higgs search modes is the search for associated production of a Higgs boson
with a W or Z vector boson, where the Z boson decays to a pair of neutrinos or the W boson decays leptonicaly and
the charged lepton is not observed and the Higgs boson decays to a pair
b quarks.  The sum of the two production mechanisms and the large branching ratios involved in the
vector boson decays make this search potentially very sensitive.
The primary background to this search is QCD dijet events where the jet energy is mis-measured
leading to apparent missing energy.
The key issue is constructing a model of the QCD background.  Both CDF and D\O~ use the comparison
of missing energy as measured by the calorimeter and the tracker to identify events with false missing
energy and build a model of the background.
The CDF and D\O~ experiments further pursue a strategy of using events where
both b jets are tagged with a loose high efficiency and low purity tag and the CDF experiment
additionally uses events where one jet is b tagged
with a tight lower efficiency higher purity tag.  
The CDF experiment uses the H1 which algorithm, which is a technique where
track information is used to measure the charged component of the jet energy and
improve the overall jet energy resolution.
Finally, CDF includes three jet events, which gives acceptance for events where the W boson decays to
tau lepton and the tau lepton decays hadronically.
The D\O~ experiment includes these events using the dedicated search described above.
For final discrimination between signal and background the experiments apply a NN algorithm in the case of CDF
and a BDT in the case of D\O.
The combination of these techniques makes this search channel the most sensitive per $fb^{-1}$ in the case of
CDF.
Using an integrated luminosity of 2.1 $fb^{-1}$ 
the $VH \rightarrow MET b\bar{b}$ searches from CDF and D\O~ have the sensitivity to see approximately 7 and 4 Higgs events
and a sensitivity of 6.3 and 8.4 times the SM production cross section for a Higgs boson of 115~GeV, respectively.

\subsection{$WH \rightarrow \ell\nu b\bar{b}$}
The most sensitive of the low mass Higgs search modes is the search for associated production of a Higgs boson
with a W boson decaying to a charged lepton and a neutrino
and the Higgs boson decaying to a pair of b quarks.
This production and decay mode enjoys the clear signal of the charged lepton
and large branching ratios for the W and Higgs boson decays.
Again the key issue is increasing lepton acceptance.  The D\O~ experiment uses it's excellent lepton
identification system to identify charge light leptons including such features as a logical OR of
all muon triggers to give full acceptance over a large range of solid angle and extended use of forward going leptons.  
CDF supplements its lepton detector coverage using leptons collected on
MET and jet triggers where the lepton is identified offline either by using areas of the detector
with muon detection systems that are not part of the trigger or identifying leptons as isolated tracks using
tracking information, which also gives acceptance for electrons and single charged hadron tau decays. 
Again both experiments apply an array
of advanced techniques to construct discriminating variables including NN, BDT and ME based discriminants.
Using an integrated luminosity of 2.7 $fb^{-1}$ 
and a combination of ME and BDT techniques the CDF experiment achieves the strongest sensitivity of any
low mass Higgs search with a sensitivity to 8 Higgs events and 5.6 
times the SM production cross section for a Higgs boson of 115~GeV.  D\O~ achieves a sensitivity to Higgs production
of 8.5 times the SM production cross section for a Higgs boson of 115~GeV. 

\subsection{$H\rightarrow W^+W^- \rightarrow \ell^+\nu\ell^-\bar{\nu}$}
The strongest sensitivity of any SM Higgs searches at the Tevatron is achieved in the
$H\rightarrow W^+W^- \rightarrow \ell^+ \nu \ell^- \bar{\nu}$ decay mode.  In this decay mode the charged leptons have
the distinct feature that they tend to be collinear due to the spin correlation of the scalar Higgs boson decaying
to vector bosons.  The CDF and D\O~ collaborations perform a search for the WW decay in the gluon fusion
and vector boson fusion production modes.  The CDF search additionally includes acceptance from associated production
of a Higgs boson with a W or Z boson.  Again the key issue is maximizing lepton acceptance and D\O~ uses 
their full lepton detection system where CDF supplements their lepton detectors with leptons identified using tracking and
evidence of a minimum ionization signature in the calorimeter.
The experiments further employ NN based discriminating variables including
kinematic information and, in the case of CDF, ME based likelihood discriminants.  The D\O~ collaboration
optimizes their analysis based on the flavor of the leptons while CDF divides the analysis in events
with 0, 1 or 2 jets.   The combination of these techniques gives the CDF and D\O~ 
experiments sensitivity to detect 17
and 16 Higgs boson events at 165~GeV, respectively, using an integrated luminosity of 3.0 $fb^{-1}$.  
The sensitivity of the experiments to set a 95\%CL limit on a Higgs boson mass
of 165~GeV are 1.6 and 1.9 times the SM production cross section.  In the absence of clear Higgs signature the
experiments set observed limits of 1.6 and 2.0 times the SM production cross section.

\subsection{SM Higgs Combination of Limits}
Substantially improved sensitivity can be achieved by combining the results of the two Tevatron experiments.
The two experiments first combine their results within their own collaborations.  At low mass this entails
combining a large number of channels.  At high mass the sensitivity is dominated by the 
$H\rightarrow W^+W^- \rightarrow \ell^+\nu\ell^-\bar{\nu}$ channel.  Both collaborations compute both individual
and combined limits including systematic uncertainties on theoretical cross sections for backgrounds and
signal, efficiencies of trigger and identification algorithms, and uncertainties that can change the shape
of the final discriminant distributions such as jet energy scale errors, QCD calculation scale variations, PDF
variations and the effects of higher order production diagrams.  The systematic uncertainties are included
as nuisance parameters and the overall fit to compute the expected and observed cross section limits can
constrain these parameters where appropriate.  The CDF collaboration uses a Bayesian technique and the D\O~
collaboration uses a CLs technique to perform the combination.  The sensitivity for the production of a Higgs
boson of 115~GeV is 3.6 and 4.6 times the SM production cross section
for the CDF and D\O~ collaborations respectively.  The
High mass results are identical to the results quoted above.

At high mass the CDF and D\O~ collaboration have combined their results between experiments.  The full combination is performed using
both the Bayesian and CLs techniques to cross check the results.   The expected sensitivity to the SM production
cross section is 1.3, 1.2, 1.4 and 1.7 times the SM cross section at masses from 160 to 175~GeV in 5~GeV increments.  Based on
the expected sensitivities at the masses
160~GeV, 165~GeV and 170~GeV the probability to exclude at lease one mass is substantial.  At 170~GeV the observed
limit excludes the SM cross section for Higgs boson production.  Similarly the second limit calculation method excludes a cross section 
5\% less than the SM cross
section for Higgs boson production.  Based on the strength of the agreement of these two methods and the good agreement with the
the expected sensitivities we report that the production of the SM Higgs boson of 170~GeV is excluded.  This exclusion
represents the first direct limits on the mass of the SM Higgs boson since the final results from the LEP
collaborations.  The full results of the high mass combination are given in~\cite{cdfd0comb} and summarized in table~\ref{limits}.
Two graphical representations showing the 95\% CL limits as a ratio to the SM cross section~\ref{limitfig} and the confidence level of the
limit~\ref{clfig} as a function of mass are shown below.

\begin{table}[ht]
\caption{\label{limits} Ratios of median expected and observed 95\% CL
limit to the SM cross section for the combined CDF and D\O~ analyses as a function
of the Higgs boson mass in ~GeV, obtained with the Bayesian method, 1, and the $CL_S$ method 2.}
\begin{tabular}{|l|c|c|c|c|c|c|c|c|c|c|c|}
\hline                     &    155  & 160 & 165 & 170 & 175 & 180 & 185 & 190 & 195 & 200\\ \hline
\hline Expected 1             &   1.7  & 1.3 & 1.2 & 1.4 & 1.7 & 2.0 & 2.8 & 3.3 & 4.2 & 4.6\\
\hline Observed 1            &   1.7  & 1.4 & 1.2 & 1.0 & 1.3 & 1.6 & 2.5 & 3.3 & 4.8 & 5.1\\
\hline Expected 2      &         1.6  & 1.2 & 1.1 & 1.3 & 1.7 & 2.0 & 2.8 & 3.4 & 4.2 & 4.7\\
\hline Observed 2      &         1.6  & 1.3 & 1.1 & 0.95& 1.2 & 1.4 & 2.3 & 3.2 & 4.7 & 5.0\\
\hline
\end{tabular}
\end{table}

\begin{figure*}[t]
\centering
\includegraphics[width=135mm]{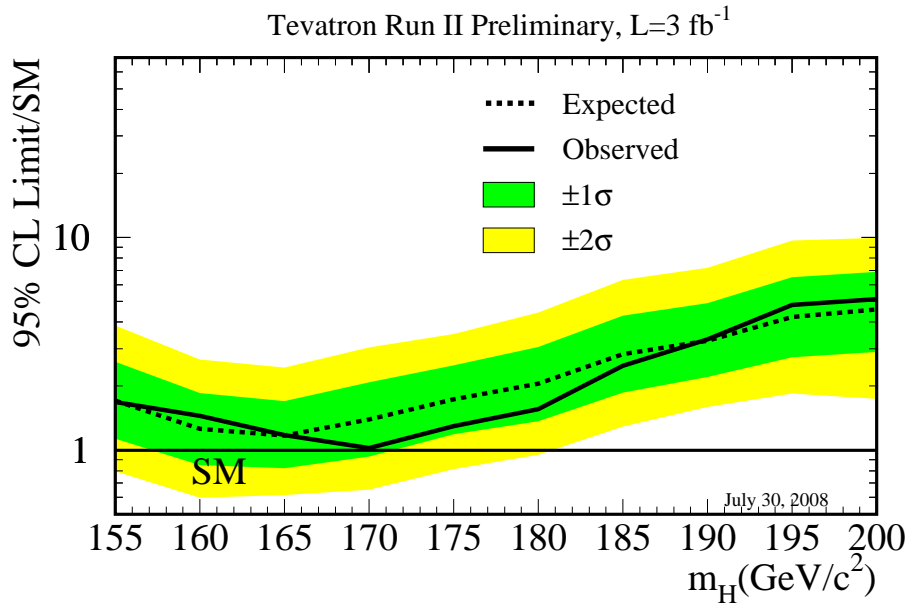}
\caption{Observed and expected (median, for the background-only hypothesis)
  95\% C.L. upper limits on the ratios to the
SM cross section,
as functions of the Higgs boson mass
for the combined CDF and D\O~ analyses.
The limits are expressed as a multiple of the SM prediction
for test masses (every 5 ~GeV/$c^2$)
for which both experiments have performed dedicated
searches in different channels.
The points are joined by straight lines
for better readability.
  The bands indicate the
68\% and 95\% probability regions where the limits can
fluctuate, in the absence of signal.} 
\label{limitfig}
\end{figure*}

\begin{figure*}[t]
\centering
\includegraphics[width=135mm]{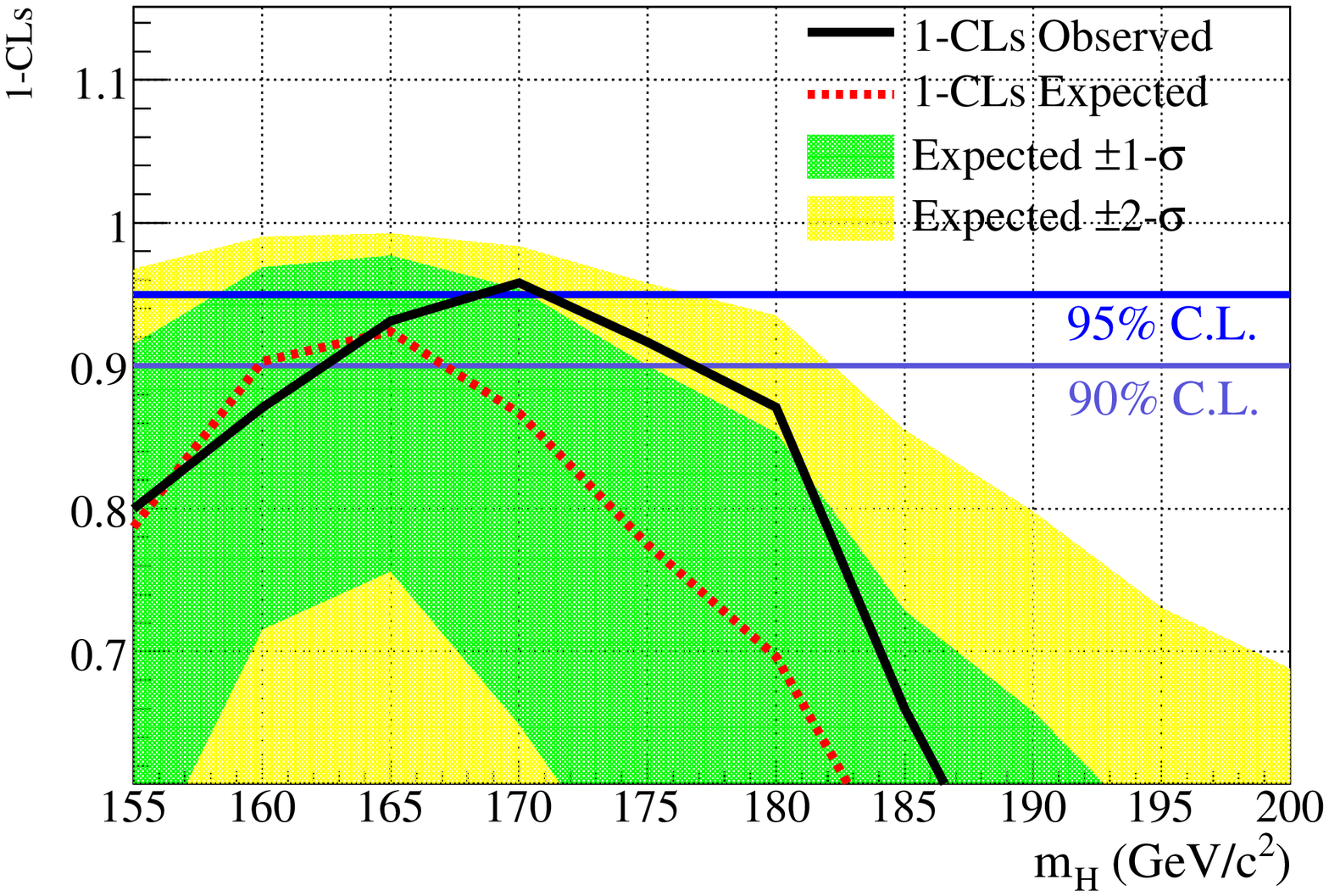}
\caption{ Distributions of 1-$CL_S$ as a function of the Higgs boson mass
(in steps of 5 ~GeV/c$^2$)
 %obtained with the $CL_s$ method
 for the combination of the
 CDF and D\O~ analyses.} 
\label{clfig}
\end{figure*}

\section{LHC Prospects}
Here I briefly comment on the LHC prospects for Higgs boson searches.  The LHC experiments have the ability to exclude
or observe the SM Higgs boson over a wide rage of masses with high significance.  At high masses above 135~GeV an
exclusion can be achieved with $1fb^{-1}$. The most difficult range will be low mass where good detector performance and
tens of $fb^-1$ will be necessary to reach SM Higgs boson sensitivity for observation.  In addition, using SUSY as a benchmark
for BSM Higgs boson searches, the LHC has sensitivity to the Higgs boson over a large region of SUSY parameters.

\section{conclusion}
I have reported on the status of Higgs boson searches and the prospects for searches at future colliders.
The Tevatron experiments have reached sensitivity for the production cross section of a SM Higgs boson at
high mass and have the potential to reach that sensitivity at all masses of interest.  Further, the Tevatron
experiments report that the production of the SM Higgs boson of 170~GeV is excluded.  With the start of the LHC
experiments we expect to have sensitivity to observe the SM Higgs boson at all masses of interest if it
exists.


\begin{thebibliography}{9}
\bibitem{notation}
We use natural units throughout in which c and hbar are taken as one and energy, momentum, and mass are referred
to in units of energy.
\end{thebibliography}
\end{document}